\documentclass[prb,twocolumn,superscriptaddress,showpacs]{revtex4}
\usepackage{epsfig}
\usepackage{color}
\usepackage{amsmath}
\usepackage[utf8]{inputenc}
\usepackage{units}

\begin{document}

\title{High energy collision cascades in tungsten: dislocation loops structure and clustering scaling laws}

\author{A.E. Sand}
\affiliation{EURATOM-Tekes, Department of Physics - P.O. Box 43, FI-00014 University of Helsinki, Finland}
\author{S.L. Dudarev}
\affiliation{EURATOM/CCFE Fusion Association, Culham Centre for Fusion Energy, Oxfordshire OX14 3DB, United Kingdom}
\author{K. Nordlund}
\affiliation{EURATOM-Tekes, Department of Physics - P.O. Box 43, FI-00014 University of Helsinki, Finland}
\date{\today}

\begin{abstract}
Recent experiments on {\it in-situ} high-energy self-ion irradiation of tungsten (W) show the occurrence of unusual cascade damage effects resulting from single ion impacts, shedding light on the nature of radiation damage expected in the tungsten components of a fusion reactor. In this paper, we investigate the dynamics of defect production in 150 keV collision cascades in W at atomic resolution, using molecular dynamics simulations and comparing predictions with experimental observations. We show that cascades in W exhibit no subcascade break-up even at high energies, producing a massive, unbroken molten area, which facilitates the formation of large defect clusters. Simulations show evidence of the formation of both $\nicefrac{1}{2}\langle111\rangle$ and $\langle100\rangle$ interstitial-type dislocation loops, as well as the occurrence of cascade collapse resulting in $\langle100\rangle$ vacancy-type dislocation loops, in excellent agreement with experimental observations. The fractal nature of the cascades gives rise to a scale-less power law type size distribution of defect clusters.
\end{abstract}


\maketitle

\section{Introduction}
Recent experiments \cite{Xiaoou} revealed unusual features of cascade damage associated with high-energy ion impacts in tungsten (W). High-energy 150 keV cascades initiated in W by self-ions were found to produce a high fraction of the $\langle100\rangle$ self-interstitial and vacancy-type dislocation loops, contradicting the current understanding of dislocation loop formation, according to which only the $\nicefrac{1}{2}\langle111\rangle$ type loops should form in tungsten \cite{Gilbert2008}, because of its elastic isotropy \cite{DBD2008}.

The highly non-equilibrium nature of cascade events implies the possibility of athermal creation of defect and dislocation structures. These processes take place over length and time scales not currently accessible to experimental methods. Atomistic simulations can aid in the understanding of defect formation mechanisms, as well as in the precise characterization of the primary state of damage produced by irradiation. The recent experiments offer a unique opportunity for a direct comparison of cascade simulations with experiment.

Tungsten has a unique combination of properties, with good thermal conductivity and high melting point, combined with resistance to erosion and low neutron activation. It is currently the preferred candidate material for the first wall components of fusion reactors \cite{Rieth2011}. However, the brittleness of tungsten poses a problem. In addition, little is known about radiation embrittlement of tungsten due to prolonged exposure to high-energy neutrons. The nature of defect structures formed in cascades constitutes the key factor determining the pathways of microstructural evolution of the material, an issue crucial to the prediction of the performance of reactor components.

The high-energy neutrons produced by the D-T fusion plasma give rise to primary recoils with energies in the hundred-keV range.
Due to the high mass of tungsten atoms, a recoil deposits its energy in a fairly compact cascade, forming a concentrated area of
high energy density. So far, atomistic simulations of damage production in W have been performed using simulation techniques based on the binary collision approximation, which are thus unable to describe the heat spike or other collective phenomena characteristic of collision cascades in metals.

Molecular dynamics (MD) simulations are capable of describing both the heat spike and the recombination of damage, and they are efficient enough to follow the complete development of cascades in systems containing millions of atoms. Cascade damage in other materials, for example iron, has been extensively studied by MD methods. In particular, it has been shown that at high impact energies cascades split into separate subcascades, resulting in defect production scaling linearly with the impact energy. For iron this occurs at energies of the order of tens of keV \cite{Stoller97}. However, in tungsten, collision cascades have previously been studied by MD methods for energies only up to 50 keV \cite{Fikar}, whereas the average energy of primary recoils from 14 MeV fusion neutrons is close to 150 keV. While the velocity of a 50 keV recoil atom in iron is 420 nm/ps, the corresponding velocity of a W recoil atom is only 230 nm/ps. Furthermore, the ratio of atomic radii of Fe and W atoms is close to 0.9, resulting in a larger cross-section for ion-ion collisions in W compared to Fe. Consequently, while Fe exhibits subcascade break-up at 50 keV \cite{Stoller97}, W does not, and therefore extrapolation to higher energies from existing data in W is not possible.

Here, we carry out and analyze MD simulations of 150 keV collision cascades in W, comparing different inter-atomic potentials. We investigate the development of the cascades, and study the effects of energy losses and the choice of primary recoil direction on the final damage, comparing simulations with observations \cite{Xiaoou}. The configurations of defect structures found here provide input for kinetic Monte Carlo and rate theory models aimed at predicting microstructural evolution in tungsten under extreme irradiation conditions.

\section{Simulation methodology}

Full collision cascades initiated by 150 keV recoils were simulated using the classical MD \cite{Allen-Tildesley} code PARCAS \cite{PARCAS}. Interatomic potentials used in the simulations include a modified EAM potential developed by Derlet {\it et al.} (D-D) \cite{Derlet}, the well-known Ackland-Thetford EAM potential (A-T) \cite{Ackland}, and a recent Tersoff-type
potential developed by Ahlgren {\it et al.} (A-H) \cite{Ahlgren}. All the potentials were smoothly fitted to the universal ZBL potential \cite{zbl} at short distances. Simulations were performed assuming the environment temperature of 0 K, in order to isolate athermal processes involved in defect production. Periodic boundary conditions were applied in all directions,
with a Berendsen thermostat \cite{Berendsen} controlling the temperature along the borders of the simulation cell.
The thermostat enables the excess heat introduced by a recoil event to be drained from the simulation cell, mimicking the dissipation that would naturally occur in the bulk of the material.

The initial momentum direction of the primary knock-on atom (PKA) was chosen randomly. To check for possible directional dependence associated with the anisotropy of the displacement threshold energy \cite{Maury}, cascades were also initiated with PKA directions taken within 2 degrees off the low index directions $\langle 100\rangle$ and $\langle 110\rangle$. The borders of the simulation cell were monitored for high energy atoms, to ensure that all the cascades were contained fully within the simulation cell. The border thermostat also serves to dampen the lower-energy pressure waves emanating from the cascade core, and inhibit their re-entry through the periodic boundaries.

Simulating multi-million atom systems requires a high degree of computational efficiency, and hence electronic interactions cannot
be directly calculated. Various methods have been developed to incorporate electronic energy losses in MD simulations of collision cascades. During the initial ballistic phase of a cascade, energy losses to electrons are well described by the available models of electronic stopping. Values for the stopping power can be fairly reliably obtained from the SRIM code \cite{SRIM}, based on theoretical analysis and experimental data. Here we employ a velocity dependent damping factor calculated from the SRIM electronic stopping power. The occurrence of a low-energy threshold for electronic stopping has been observed in semiconductors \cite{Dra05,Pru07}, and is explained by the band gap inhibiting excitation of electrons below a certain energy. Similar effects have also been reported in metals \cite{Valdes94}. While the occurrence of a threshold in tungsten lacks direct evidence, introducing a cut-off energy when employing a velocity dependent damping is necessary to avoid quenching all the thermal modes.

Le Page {\it et al.} assessed the accuracy of several semi-classical models for energy losses during the ballistic phase of cascades in Cu with PKA energies $E_{PKA}\leq 1$ keV, by comparing to
the quantum-mechanical Ehrenfest approximation \cite{lePage}. They showed that when a damping term with a cut-off is employed, a cut-off energy of $T_c \approx 1$ eV is optimal. Although somewhat less precise than other models, the accuracy of this simple approximation increases for higher PKA energies, although some dependence of results on the choice of $T_c$ is retained.

We employ a range of different cut-off values, in order to identify possible spurious effects associated with the choice of $T_c$, for which the literature data vary and which is often chosen arbitrarily. We also compare the above approach to a method used in earlier cascade simulations \cite{Stoller97}, where the energy lost to electrons is
subtracted prior to a simulation, and the initial PKA is assumed to have the energy corresponding to the total damage energy, {\it i.e.} the difference between the recoil energy and the energy lost to electrons.

\section{Results}

The size and shape of a cascade were determined by analysing the liquid atom area at the point when this area is at its largest, approximately after the first 250 ps. At this point the typical extent of the liquid area was between 150 and 250 Angstroms.
An atom was determined to belong to the liquid phase using a kinetic energy criterion which provides a good approximation to a more sophisticated structure factor analysis \cite{Nor98}. The extent of the liquid area correlates well with the distribution of the final damage, with larger self-interstitial atom (SIA) clusters wrapped around the liquid area, and single SIAs located outside this area. Vacancies are concentrated towards the core of the melt.

A visual inspection of the development of the cascade showed that the liquid area was often strongly elongated, indicating that the cascade energy was close to the threshold for subcascade break-up, but in the simulations performed here, totaling almost 100, no subcascades were observed. The liquid area, although irregular, was unbroken in all cases, forming a large region which at the peak of its size contained roughly 30000 atoms with an average energy density of almost 4 eV per atom.

\begin{figure}
\includegraphics[width=\columnwidth]{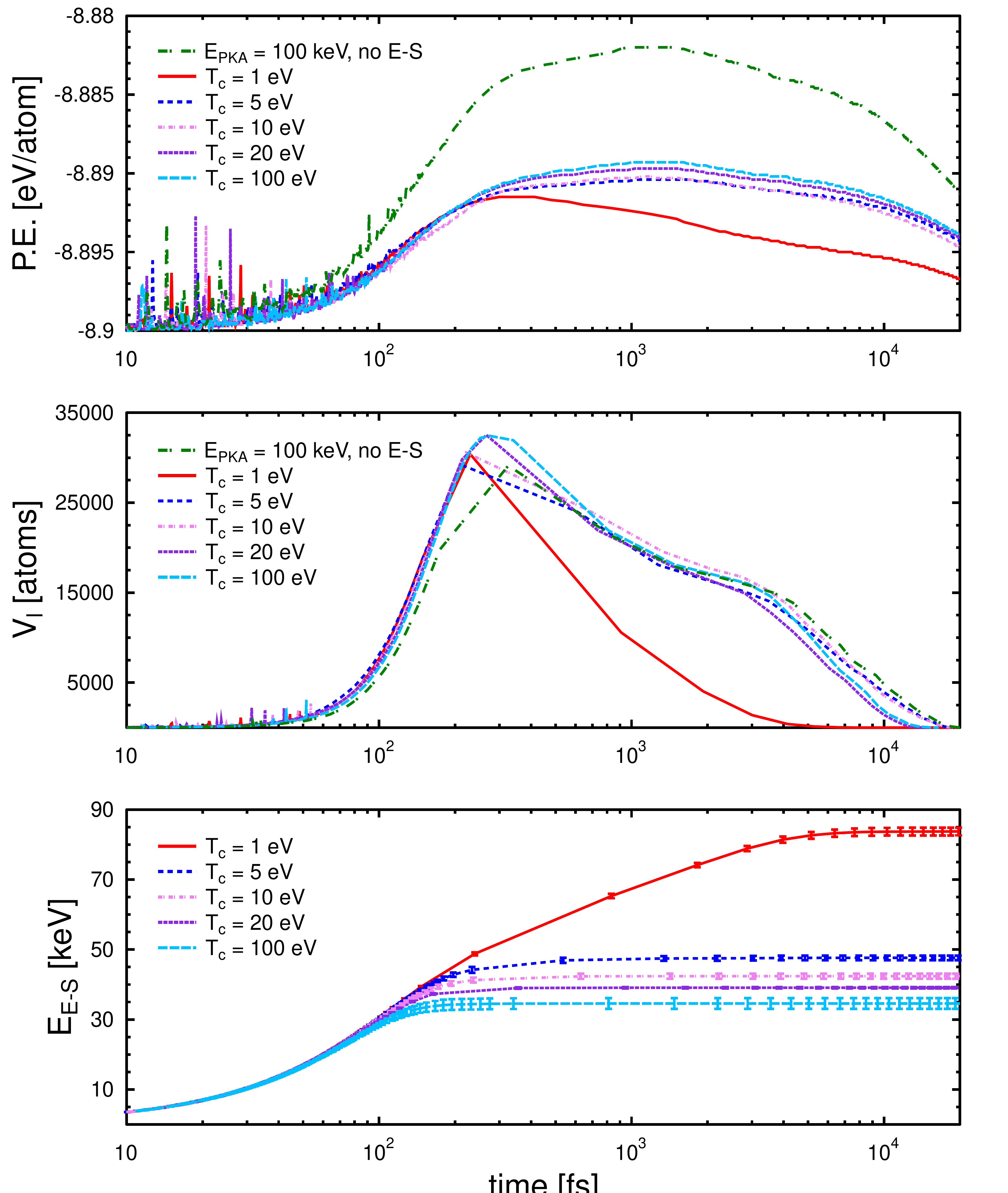}
\caption{(Color online) Time evolution of cascades simulated using the D-D potential for several different electronic stopping (E-S) cut-off values $T_c$. (a) Top: potential energy per atom for representative cascades. (b) Middle: volume of the liquid area given as the number of liquid atoms for representative cascades. (c) Bottom: the total energy lost to electronic stopping, averaged over the cascades performed for each cut-off value.}
\label{fig.1}
\end{figure}

The plot of energy losses to electrons (fig. \ref{fig.1} (c)) shows no difference during the first 100 fs for different $T_c$, due to the negligible number of atoms with energies comparable to the cut-off energy at this time. This constitutes the ballistic phase of the cascade, which is characterized by a sharp increase in the potential energy (see fig. \ref{fig.1} (a)), and fast expansion of the strongly disordered liquid-like area (fig. \ref{fig.1} (b)). In other words, energetic ions are penetrating the undisturbed areas of the lattice, and electronic stopping models are physically motivated.  After around 200 fs the ions have ceased to travel
individually and form an increasingly uniform front, which subsequently transforms into a pressure wave causing only elastic displacements, ultimately leaving the lattice intact. This marks the transition to the thermal phase of the cascade.
The liquid area thermalizes and the temperature in the system rises slightly, while the rate of potential energy increase slows considerably. These two different phases of development, occurring before and after the number of liquid atoms reaches its peak,
correspond well to the ``destructive'' and ``non-destructive'' phases of initial cascade development found by Calder \cite{Calder2010}. During the transition, atoms in the energy range 1 - 100 eV come into play, and at this point the choice of $T_c$ begins to have an impact. However, the non-destructive phase no longer involves single ions penetrating the undisturbed regular atomic lattice, and thus theories of electronic stopping are no longer directly applicable. For $T_c \geq 5$ eV, no further energy is lost to the electronic system during the thermal phase. However, when $T_c=1$ eV, an additional 40 keV is lost during this non-destructive phase, effectively from the random motion of the hot atoms in the molten cascade core, where the thermal energy is sufficiently high so that the kinetic energy of many atoms lies between 1-5 eV. This energy loss causes the cascade core to be cooled down very quickly compared to the simulations assuming higher $T_c$. As long as the electronic stopping doesn't affect atoms in the melt during the non-destructive phase, the effect of varying $T_c$ is minor, even up to 100 eV.

To further investigate the effect of time dependence of energy losses, cascades were simulated with an initial PKA energy of 100 keV and no electronic stopping. As seen from the plot in fig. \ref{fig.1} (a), the potential energy rises almost twice as much during the 100 keV cascade as it does in a simulation with electronic stopping, despite the fact that the damage energy in the latter case is close to 110 keV. Since the size of the liquid area is more or less the same in both cases, the extra potential energy in the 100 keV cascades is distributed among the same number of atoms, indicating strong lattice distortion. Correspondingly, the largest defect structures were produced in such cascades. This shows that the mechanism and dynamics of energy losses do affect the final defect configurations, and thus a dynamic treatment of electronic energy losses is necessary to ensure that simulations are realistic.

Defect numbers were analyzed using an automated Wigner-Seitz (W-S) method \cite{Nor97}. The results are compared visually to the energy filtered atomic configurations, and an exact match was established in all cases. In bcc metals the W-S method gives a well defined value characterizing the actual damage, as a direct count of the number of SIAs and vacancies present in the lattice.

Cascades extended in random directions, uncorrelated with the initial direction of the PKA, which thus had little effect on the final damage. Large SIA clusters were slightly more frequent with recoils in a $\langle 110 \rangle$-type direction, but the difference in clustering statistics was within the statistical error. Because of large differences between the individual cascades, any given single cascade may not be representative of damage production in general, and thus good statistics were necessary to determine the accurate defect production rates.

\begin{table*}
\caption{Average number of defects $N_{SIA}$, the size of the largest SIA cluster $N_{cl}^{max}$, and the fraction of vacancies ($F_{vac}$) and SIAs ($F_{SIA}$) in clusters of 4 or more, for various simulation parameters. Cascades with $E_{PKA}=100$ keV did not include further electronic stopping. The final column shows the number of simulations performed for a given set of parameters.}
\label{table2}
\begin{center}
\begin{tabular}{lcccccr}
Potential & $T_c$  & $N_{SIA}$ & $N_{cl}^{max}$ & $F_{vac}$  & $F_{SIA}$ & $N_{sim}$  \\
\hline
D-D & 1 eV   & 122 $\pm$ 2.9  & 54  & 0.00         & 0.47 $\pm$ 0.03    & 10 \\
D-D & 5 eV  & 183  $\pm$   27.9 & 164 & 0.21 $\pm$   0.07  &  0.71  $\pm$  0.07  & 9  \\
D-D & 10 eV  & 179 $\pm$  21.0  & 175   & 0.19 $\pm$ 0.06   & 0.72 $\pm$ 0.06    & 10 \\
D-D & 10 eV; $\langle 100\rangle$  & 184  $\pm$   27.0 & 130  & 0.17  $\pm$  0.05 & 0.74  $\pm$  0.06 & 9 \\
D-D & 10 eV; $ \langle 110\rangle$  & 202  $\pm$   31.8 & 233  & 0.18  $\pm$  0.08 & 0.74  $\pm$  0.05 & 9 \\
D-D & 20 eV   &  183  $\pm$   18.1 & 94  &  0.16  $\pm$  0.06 & 0.81  $\pm$  0.01 & 5 \\
D-D & 100 eV   & 257  $\pm$   49.7  & 224 & 0.28  $\pm$  0.10 & 0.81  $\pm$  0.05 & 5 \\
D-D & $E_{PKA}=100$ keV  & 250 $\pm$  51.4  & 433   & 0.32 $\pm$ 0.09   & 0.78 $\pm$ 0.06    & 10  \\
A-T & 1 eV & 93  $\pm$   4.3 & 16  & 0.04  $\pm$  0.01 & 0.34 $\pm$   0.03 & 10 \\
A-T & 10 eV & 124  $\pm$   16.2 & 85 & 0.35  $\pm$  0.08 & 0.66  $\pm$  0.06 & 10 \\
A-H & 10 eV & 92  $\pm$   8.5 & 41 & 0.05  $\pm$  0.03 & 0.41  $\pm$  0.08 & 5 \\
\end{tabular}
\end{center}
\end{table*}

Defect numbers and clustering statistics for various cascade parameters are given in Table \ref{table2}. A vacancy (or a SIA) was determined as belonging to a cluster if it was within the 2nd (3rd) nearest neighbor distance of another vacancy (SIA) in the cluster. The lowest average number of defects were produced in simulations with $T_c=1$ eV, with 120 Frenkel pairs per ion. Higher cut-off values all yielded around 200 Frenkel pairs per ion. However, the largest effect of the cut-off value was seen in the amount of clustering of vacancies. With  $T_c=1$ eV, no vacancy clusters formed. This is not compatible with experimental results, where
visible vacancy-type dislocation loops \cite{Xiaoou} or depleted zones \cite{Pramanik83} are seen to occasionally result from single ion impacts. Therefore simulations with low cut-offs do not represent the actual cascade evolution. It is interesting to note that larger defect numbers, seen in simulations with higher cut-off energy, also show higher statistical fluctuations. This is because cascades with these parameters nevertheless sometimes produce very few fairly small clusters. In those cases the final defect numbers and damage configurations were similar to the cascades with $T_c=1$ eV. On the other hand, when large clusters did form, defect numbers were also high. This is in agreement with findings by Calder {\it et al.} \cite{Calder2010}, who also noted that large clusters were statistically correlated with large defect numbers. The A-T potential showed the same trend as the D-D potential, with very little clustering occurring for $T_c=1$ eV. With $T_c=10$ eV, both SIAs and vacancies formed larger clusters, although the individual cluster sizes were smaller than those predicted using the D-D potential. The Tersoff-type potential showed very little clustering even for $T_c=10$ eV, producing only one cluster containing in excess of 40 SIAs.

\begin{figure}
\includegraphics[width=\columnwidth]{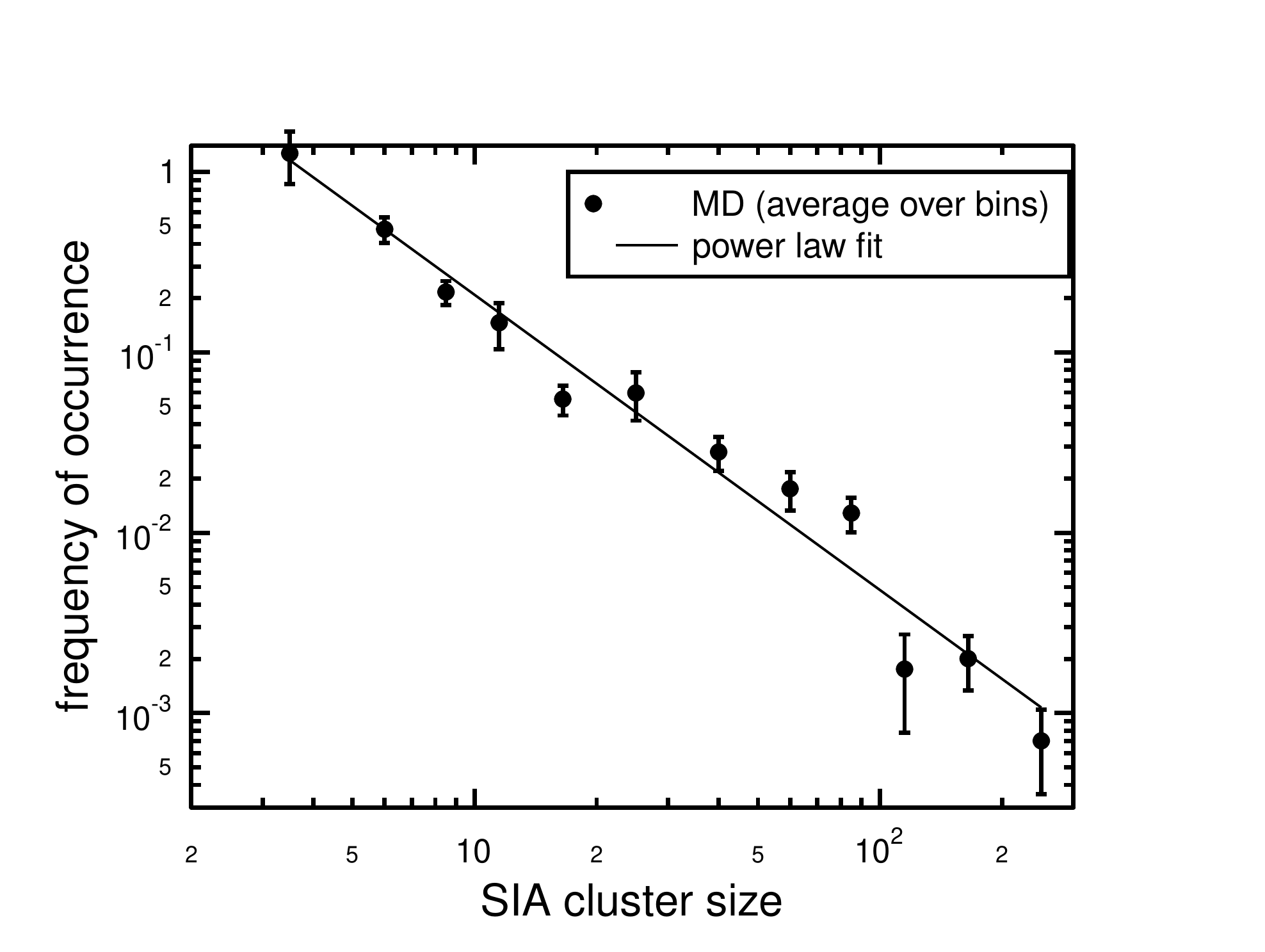}
\caption{Frequency-size distribution of clusters with three or more SIAs, from simulations performed using the D-D potential and $T_c \geq 5$ eV. The data are averaged over bins and fitted to a power law function.}
\label{fig.3}
\end{figure}

\begin{figure}
\begin{center}
\includegraphics[width=0.9\columnwidth]{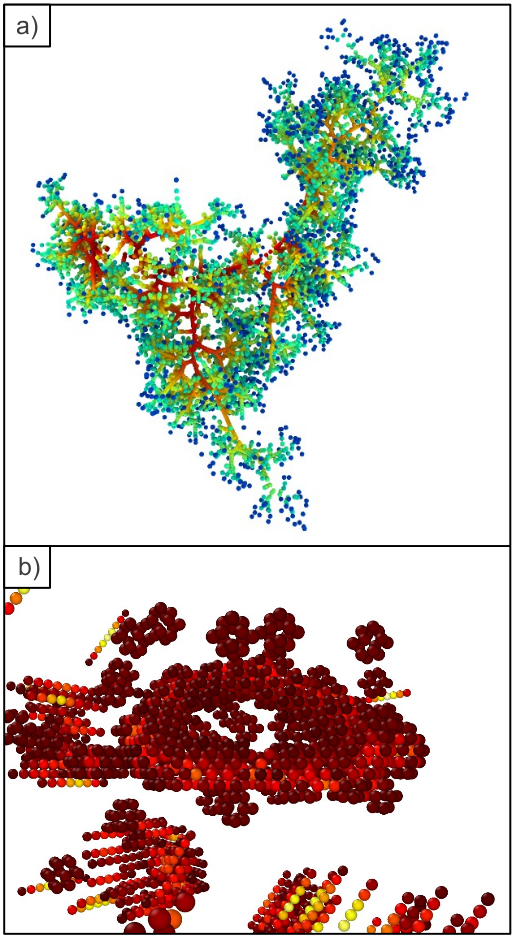}
\caption{(Color online) a) Paths of energetic ions ($\geq 10 eV$) during the initial development of the cascade.
The color scale indicates the time, starting with red at 0 fs and ending with blue at 200 fs. b) Final configuration from a cascade simulation resulting in cascade collapse, showing a large dislocation loop with Burgers vector $b=\langle100\rangle$. Atoms are colored according to their potential energy, with lighter color representing higher energy.}
\label{fractal}
\end{center}
\end{figure}

No exponential function was found to fit the frequency-size distribution of SIA clusters well. Rather, the data closely follow a power law shown in fig. \ref{fig.3} and given by the equation
\begin{equation}
\label{eq.1}
F(n)={A\over n^S}
\end{equation}
where
\begin{equation}
\label{eq.2}
A=7.45 \pm 1.52 \text{\ \ and\ \ } S=1.63 \pm 0.07
\end{equation}
The data for the plot shown in fig. \ref{fig.3} were gathered from all the cascades simulated in this work using the D-D potential and $T_c \geq 5$ eV, and were sorted into logarithmic bins. Different bin sizes were tested and found to have a minimal effect on the values of parameters $A$ and $S$. Power laws similar to that given by eq. (\ref{eq.1}) are often found in self-organized critical systems, such as the sandpile model by Bak {\it et al.} \cite{Bak88} or in fractal systems such as forest fires and landslides \cite{Turcotte}. The fractal nature of cascade development \cite{Mar90} is illustrated in fig. \ref{fractal}. Fractal systems are well known to exhibit power law behaviour and are inherently scale invariant. The emergence of a power law distribution for defect cluster sizes thus shows that there is no characteristic energy scale determining an average cluster size produced in a cascade. Similar lack of energy dependence has also been seen in MD simulations of cascades up to 100 keV in iron \cite{Souidi11}. 
Since the damage energy available in a dense cascade region is orders of magnitude larger than that stored in the resultant defects, in principle almost arbitrarily large defects may form, within the range of validity of the power law. 
Despite the relative rarity of occurrence of large defect clusters found in this study, the available data give no indication that the distribution deviates from the power law for cluster sizes up to a few hundred SIAs. However, a natural upper bound to the cluster size is imposed by the size of the
melt, since large clusters are formed within or in the immediate vicinity of this area, and only single SIAs are found further out. Assuming the power law is valid to $n^*\approx 600$, and that the distribution swiftly approaches zero after that, the total average number of Frenkel pairs produced in a cascade is approximately given by $\int_{1}^{n^*}nF(n)dn\approx 200$, which agrees well with the numbers obtained in these simulations.
The clustering scaling law also provides a convenient means for estimating the relative probability of forming large defect clusters. For example, assuming that for $n\gg1$ the same power law applies to vacancies, and noting that a cascade produces identical numbers of vacancies and SIAs, we find that the probability of forming a loop with diameter greater than $\sim 3$nm (containing in excess of $\sim 150$ defects \cite{Gilbert2008}) approximately equals $\int _{150}^{n^*}F(n)dn/ \int_{1}^{n^*}F(n)dn\approx 3\%$, in good agreement with experimental observations \cite{Xiaoou}. This estimate illustrates the predictive power of a scale-invariant power law (\ref{eq.1}), which makes it possible to circumvent the need to simulate a large number of cascades directly.

Each large defect cluster was analyzed separately to determine its dislocation loop character. In D-D potential simulations, approximately half of all the larger clusters formed complex configurations, although often partial loop structure was evident.
Such defects, if stable, would be immobile, but for a number of cases the defect relaxed to a $b=\nicefrac{1}{2}\langle111\rangle$
dislocation loop after 100 ps at 600 K. Clearly defined loops with more than 80 SIAs were all found to have the Burgers vector $b=\nicefrac{1}{2}\langle111\rangle$, while almost half of the smaller loops with 35-80 SIAs had Burgers vector $b=\langle100\rangle$ whereas the other half had $b=\nicefrac{1}{2}\langle111\rangle$. Smaller clusters were composed of parallel crowdions.
In all the cases where well formed dislocation loops were identified, their habit plane normal and their Burgers vector were not aligned. With the A-T potential, SIA clusters of 30 or more formed clear dislocation loops, most of them with Burgers vector
$b=\nicefrac{1}{2}\langle111\rangle$, although one $b=\langle100\rangle$ loop was seen.

Vacancy clusters formed mostly as low density areas at the center of what had been the liquid area. No vacancy dislocation loops were found in the D-D potential simulations, where the loop stability criterion required that a closed loop had to contain more than $\sim 150$ vacancies \cite{Gilbert2008} to form. The probability of cascade collapse was found to be affected by the rate of cooling of
the heat spike, with slower cooling increasing the probability of collapse \cite{English87}. The clustered vacancies observed in simulations may therefore indicate cascades close to collapsing, which may collapse in higher temperature simulations.
With the A-T potential, on the other hand, one cascade collapsed into a vacancy loop, shown in fig. \ref{fractal}, composed of 213 vacancies, with Burgers vector $b=\langle100\rangle$. Vacancy clustering was more sensitive than SIA clustering to the choice of $T_c$, with greater clustering occurring for higher $T_c$. This is due to the slower rate of cooling of the liquid area, which allows the vacancies to be pushed towards the cascade center as the recrystallization front progresses inwards \cite{Nor98d}. Since quenching of the thermal spike is more rapid at 0 K than at finite temperatures, the effect of $T_c$ on defect clustering may be additionally enhanced in these simulations.

\section{Conclusions}

The MD simulations described in this work show the formation of $b=\langle100\rangle$ as well as $b=1/2\langle111\rangle$ dislocation loops in tungsten as a result of irradiation-induced collision cascades, in excellent agreement with recent experimental findings \cite{Xiaoou}. Both vacancy- and SIA-type dislocation loops are found, with the occurrence of SIA-type loops being more probable than cascade collapse into a vacancy-type loop.

Two different EAM type potentials predicted similar behaviour with regard to the SIA-type dislocation loops, whereas the probability of cascade collapse was more sensitive to the choice of the potential. The Tersoff-type potential tested in this work did not reproduce experimental findings in relation to the formation of visible large defect structures in cascades. The choice of the cut-off energy $T_c$ for electronic stopping was found to have a clear effect on defect clustering. A low cut-off of 1 eV inhibited the
formation of large defect structures, whereas $T_c \geq 5$ eV resulted in the occasional formation of large SIA and vacancy clusters.

High-energy cascades in tungsten exhibited no splitting into subcascades, in agreement with the earlier work exploring a lower energy range. The frequency-size distribution of SIA clusters was found to closely follow a power law, highlighting the scale-invariant nature of defect clustering and formation of dislocation loops.

\section*{Acknowledgements}
This work, supported by the European Communities under the contract of Association between EURATOM/Tekes, was carried out within the framework of the European Fusion Development Agreement. Partial support was also received from the EURATOM 7th framework programme, under grant agreement number 212175 (GetMat project). Work at CCFE was funded by the RCUK Energy Programme under grant EP/I501045 and the European Communities under the contract of association between EURATOM and CCFE. The views and opinions expressed herein do not necessarily reflect those of the European Commission. Grants for computer time from the Centre for Scientific Computing in Espoo, Finland, and funding from EURATOM staff mobility programme are gratefully acknowledged.


\begin{thebibliography}{0}

\bibitem{Xiaoou}
Yi X., Jenkins M. L., Brice\~no M., Roberts S. G., Zhou Z., Kirk M. A.,
Phil. Mag (2013) DOI:10.1080/14786435.2012.754110

\bibitem{Gilbert2008}
Gilbert M. R., Dudarev S. L., Derlet P. M., Pettifor D. G.,
J. Phys.: Condens. Matter 20 (2008) 345214

\bibitem{DBD2008}
Dudarev S. L., Bullough R., Derlet P. M.,
Phys. Rev. Lett. 100 (2008) 135503

\bibitem{Rieth2011}
Rieth M. {\it et al.},
J. Nucl. Mat. 417 (2011) 463-467X

\bibitem{Stoller97}
Stoller R. E., Odette G. R., Wirth B. D.,
J. Nucl. Mat. 251 (1997) 49-60

\bibitem{Fikar}
Fikar J., Schäublin R.,
J. Nucl. Mat. 386-388 (2009) 97-101

\bibitem{Allen-Tildesley}
Allen M.P., Tildesley D.J.,
  \emph{Computer Simulation of Liquids},
Oxford University Press, Oxford
(1989)

\bibitem{PARCAS}
Nordlund K.,
{\sc parcas} computer code,
(2006)

\bibitem{Derlet}
Derlet P. M., Nguyen-Manh D., Dudarev S. L.,
Phys. Rev. B 76 (2007) 054107

\bibitem{Ackland}
Ackland G. J.,  Thetford R.,
Phil. Mag. A 56 (1987) 15-30

\bibitem{Ahlgren}
Ahlgren T., Heinola K., Juslin N., Kuronen A.,
J. Appl. Phys. 107 (2010) 033516

\bibitem{zbl}
Ziegler J. F., Biersack J. P., Littmark U.,
  \emph{The Stopping and Range of Ions in Matter},
Pergamon, New York
(1985)

\bibitem{Berendsen}
Berendsen H. J. C., Postma J. P. M., van Gunsteren W. F., DiNola A., Haak J. R.,
J. Chem. Phys. 81 (1984) 3684

\bibitem{Maury}
Maury F., Biget M., Vajda P., Lucasson A., Lucasson P.,
Radiation effects 38 (1978) 53-65

\bibitem{SRIM}
Ziegler J. F.,
  \emph{SRIM software package} available online at http://www.srim.org

\bibitem{Dra05}
Draxler M., Chenakin S. P., Markin S. N., Bauer P.,
Phys. Rev. Lett. 95 (2005) 113201

\bibitem{Pru07}
Pruneda J. M., S\'anchez-Portal D., Arnau A., Juaristi J. I., Artacho E.,
Phys. Rev. Lett. 99 (2007) 235501

\bibitem{Valdes94}
Vald\'{e}s J. E., Eckardt J. C., Lantschner G. H., Arista N. R.,
Phys. Rev. A 49 (1994) 1083

\bibitem{lePage}
le Page J., Mason D. R., Race C. P., Foulkes W. M. C.,
New Journal of Physics 11 (2009) 013004

\bibitem{Nor98}
Nordlund K., Ghaly M., Averback R. S., Caturla M., Diaz de la Rubia T., Tarus J.,
Phys. Rev. B 57 (1998) 7556

\bibitem{Calder2010}
Calder A. F., Bacon D. J., Barashev A. V., Osetsky Y. N.
Phil. Mag. 90 (2010) 863

\bibitem{Nor97}
Nordlund K., Averback R. S.,
Phys. Rev. B 56 (1997) 2421

\bibitem{Pramanik83}
Pramanik D., Seidman D. N.,
J. Appl. Phys. 54 (1983) 6352

\bibitem{Bak88}
Bak P., Tang C., Wiesenfeld K.,
Phys. Rev. A 38 (1987) 364

\bibitem{Turcotte}
Turcotte B. L.,
Rep. Prog. Phys. 62 (1999) 1377

\bibitem{Mar90}
Moreno-Marin J. C., Conrad U., Urbassek H. M., Gras-Marti A.
Nucl. Instr. and Meth. B 48 (1990) 404

\bibitem{Souidi11}
Souidi A., Hou M., Becquart C. S., Malerba L., Domain C., Stoller R. E., 
J. Nucl. Mat. 419 (2011) 122-133

\bibitem{English87}
English C. A., Jenkins M. L.,
Materials Science Forum 15-18 (1987) 1003

\bibitem{Nor98d}
Nordlund K., Averback R. S.,
Phys. Rev. B 59 (1999) 20-23

\end{thebibliography}
\end{document}